\documentclass[preprint,showpacs,preprintnumbers,amsmath,amssymb]{revtex4}

\usepackage{graphicx,subfigure}
\usepackage{dcolumn}
\usepackage{bm}
\usepackage{amsmath}

\newcommand{\be}{\begin{equation}}
\newcommand{\ee}{\end{equation}}
\newcommand{\bea}{\begin{eqnarray}}
\newcommand{\eea}{\end{eqnarray}}

\newcommand{\GeV}{\rm GeV}

\begin{document}

\bibliographystyle{apsrev}

\preprint{UAB-FT-561}

\title{Planck-Scale Effects on Global Symmetries: \\
Cosmology of Pseudo-Goldstone Bosons
} 

\author{Eduard Mass{\'o}}

\author{Francesc Rota}

\author{Gabriel Zsembinszki}

\affiliation{Grup de F{\'\i}sica Te{\`o}rica and Institut
de F{\'\i}sica d'Altes
Energies\\Universitat Aut{\`o}noma de Barcelona\\
08193 Bellaterra, Barcelona, Spain}


\date{\today}

\begin{abstract}
We consider a model with a small explicit breaking of a global
symmetry, as suggested by gravitational arguments. Our model has
one scalar field transforming under a non-anomalous $U(1)$ symmetry, and coupled to matter and to gauge bosons. The
spontaneous breaking of the explicitly broken symmetry gives rise to a massive
pseudo-Goldstone boson. We analyze thermal and non-thermal
production of this particle in the early universe, and perform a
systematic study of astrophysical and cosmological constraints on
its properties. We find that for very suppressed explicit breaking
the pseudo-Goldstone boson is a cold dark matter candidate.

\end{abstract}

\maketitle


\section{Introduction}
\label{introduction} It is generally believed that there is new
physics beyond the standard model of particle physics. At higher
energies, new structures should become observable. Among them, there will probably be new global
symmetries that are not manifest at low energies. It is usually
assumed that symmetries would be restored at the high temperatures
and densities of the early universe.

However, the restoration of global symmetries might be not
completely exact, since Planck-scale physics is believed to break
them explicitly. This feature comes from the fact that black holes
do not have defined global charges and, consequently, in a
scattering process with black holes, global charges of the
symmetry would not be conserved \cite{banks}. Wormholes provide
explicit mechanisms of such non-conservation \cite{Giddings_etal}.

In the present article, we are concerned with the case that the
high temperature phase is only approximately symmetric. The
breaking will be explicit, albeit small. In the process of
spontaneous symmetry breaking (SSB), pseudo-Goldstone bosons
(PGBs) with a small mass appear. We explore the cosmological
consequences of such particle species, in a simple model that
exhibits the main physical features we would like to study.

The model has a (complex) scalar field $\Psi(x)$
transforming under a global, non-anomalous,  $U(1)$ symmetry. We do not need
to specify which quantum number generates the symmetry; it might
be B-L, or a family $U(1)$ symmetry, etc. We assume that the
potential energy for $\Psi$ has a symmetric term and a symmetry-breaking term
\begin{equation}
V=V_{sym}+V_{non-sym}
\label{V}
\end{equation}
The symmetric part of the potential is
\begin{equation}
V_{sym}=\frac14\lambda[|\Psi|^2-v^2]^2
\label{Vsym}
\end{equation}
where $\lambda$ is a coupling and $v$ is the energy scale of the
SSB. This part of the potential, as well as the kinetic term
$|\partial^{\mu} \Psi|^2$, are invariant under the $U(1)$ global
transformation $\Psi\rightarrow e^{i\alpha}\Psi$.

Without any clue about the
precise mechanism that generates $V_{non-sym}$, we work in an
effective theory framework, where operators of order
higher than four break explicitly the global symmetry. The
operators would be generated at the Planck scale $M_P=1.2 \times
10^{19}$ GeV and are to be used at energies below
$M_P$. They are multiplied by inverse powers of $M_P$, so that when $M_P\rightarrow\infty$ the new
effects vanish.

The $U(1)$ global symmetry is preserved by $\Psi^{\star}\Psi=|\Psi|^2$
but is violated by a single factor $\Psi$. So, the simplest new
operator will contain a factor $\Psi$
\begin{equation}
V_{non-sym}=-g\frac1{M_P^{n-3}}|\Psi|^n\left(\Psi e^{-i\delta}+\Psi^{\star}
e^{i\delta}\right)
\label{Vnosym}
\end{equation}
with an integer $n\geq4$. The coupling in (\ref{Vnosym}) is in
principle complex, so that we write it as $g\, e^{-i\delta}$ with
$g$ real. We will consider that $V_{non-sym}$ is small enough so
that it may be considered as a perturbation of $V_{sym}$. Even if
(\ref{Vnosym}) is already suppressed by powers of the small factor
$v/M_P$, we will assume $g$ small. In fact, after our
phenomenological study we will see that $g$ must be tiny.

To study the modifications that the small explicit symmetry
breaking term induces in the SSB process, we use
\begin{equation}
\Psi=(\rho+v)e^{i\theta/v}
\label{Psiparam}
\end{equation}
with new real fields $\rho(x)$ and the PGB $\theta(x)$. Introducing (\ref{Psiparam}) in (\ref{Vnosym}) we get
\begin{equation}
V_{non-sym}=-2\,g\,v^4\left(\frac{v}{M_P}\right)^{n-3}
\cos{\left(\frac{\theta}{v}-\delta\right)}+\cdots
\label{Vnonsym2}
\end{equation}
The dots refer to terms where $\rho(x)$ is present. We see from
(\ref{Vnonsym2}) that there is a unique vacuum state, with
$<\theta>=\delta\, v$. To simplify, we
redefine  $\theta' = \theta- \delta\, v$, and drop the prime, so
that the minimum is now at $<\theta>=0$.  From (\ref{Vnonsym2}) we easily obtain the $\theta$ particle mass
\begin{equation}
m_{\theta}^2=2g\left(\frac{v}{M_P}\right)^{n-1}M_P^2
\label{thetamass}
\end{equation}

Although for the sake of generality we keep the $n$-dependence in
(\ref{thetamass}), when discussing numerics and in the figures we
particularize to the simplest case $n=4$, with the operator in
(\ref{Vnosym}) of dimension five. We will discuss in Section
\ref{conclusions} what happens for $n>4$.

To fully specify our model, we finally write the couplings of the
PGB $\theta$  to other particles. We have the usual derivative
couplings to fermions \cite{gelmini},
\begin{equation}
{\cal L}_{\theta f\bar f} = \frac{g'}{2\,v}\, (\partial_\mu \theta)\,
 \bar f \gamma^\mu \gamma_5 f = g_{\theta f \bar
f}\, \theta\, \bar f \gamma_5 f
\label{ferm_coup}
\end{equation}
We have no reason to make $g'$ very different from
${\cal O}(1)$. To have less parameters
we set $g'=1$ for all fermions and discuss in a final
section about this assumption. Then
\begin{equation}
  g_{\theta f\bar f} = \frac{m_f}{v}
\end{equation}

The PGB $\theta$ couples to two photons through a loop,
\begin{equation}
  {\cal
L}_{\theta \gamma \gamma} = \frac{1}{8}\, g_{\theta \gamma
\gamma}\, \epsilon^{\mu\nu\rho\tau} F_{\mu\nu}F_{\rho\tau} \,
\theta
\label{phot_coup}
\end{equation}
where the effective coupling is $g_{\theta \gamma \gamma}
= \frac{8\alpha}{\pi v}$. In the same way, there is a coupling to
two gluons, ${\cal L}_{\theta gg} = \frac{1}{8}\, g_{\theta gg}\,
\epsilon^{\mu\nu\rho\tau} G^a_{\mu\nu}G^a_{\rho\tau}\, \theta$
with $g_{\theta gg } = \frac{3\alpha_s}{\pi v}$. Couplings of
$\theta$ to the weak gauge bosons do not play a relevant role in
our study.

We have now the model defined. It has two free parameters: $g$,
which is the strength of the explicit symmetry breaking term in
(\ref{Vnosym}), and $v$, which is the energy scale of the SSB
appearing in (\ref{Vsym}).

At this point we would like to comment on the similarities and
differences between our $\theta$-particle and the axion
\cite{Weinberg:1977ma}. There are similitudes because of their
alike origin. For example the form of the coupling
(\ref{ferm_coup}) to matter and (\ref{phot_coup}) to photons is
entirely analogous. A consequence is that we can borrow the
supernova constraints on axions \cite{Raffelt_book} to constrain
our PGB. Another analogy is that axions are produced in the early
universe by the misalignment mechanism \cite{three_papers} and by
string decay \cite{vilenkin} and $\theta$-particles can also arise
in both ways.

However, the fact that the $\theta$-particle gets its mass just
after the SSB while axions become massive at the QCD scale
introduces important differences. While in the axion case clearly
only the angular oscillations matter and always one ends up with
axion creation, in our model we need to work in detail the coupled
evolution equations of the radial and angular part. We will need
to establish when there are and when there are not angular
oscillations leading to PGB creation.

Also, from the phenomenological point of view, the axion model is
a one-parameter model while our model has two parameters $v$ and
$g$. As we will see, this extra freedom makes the supernova
constraints to allow a relatively massive $\theta$-particle, and
we will need to investigate the consequences of the decay in the
early universe. There is no analogous study for axions, simply
because the invisible axion is stable, in practical terms.

Let us also mention other previous work on PGBs. In \cite{Mohapatra}, the
authors consider the SSB of B-L with explicit gravitational
breaking so that they obtain a massive majoron. An important
difference between their work and ours is that we consider
non-thermal production, which is
crucial for our conclusions about the PGB being a dark matter
candidate. Also, there is previous work on explicit breaking of
global symmetries \cite{Ross},
and specifically on Planck-scale breaking \cite{Lusignoli}.
Cosmological consequences of some classes of PGBs are discussed in \cite{Hill}.
Finally, let us mention a recent paper \cite{Kazanas:2004kv} where a massive
majoron
is considered in the context of a supersymmetric singlet majoron
model \cite{ref3}. An objective of the authors of
\cite{Kazanas:2004kv} is to get hybrid inflation. Compared to our
work, they consider other phenomenological consequences that the
ones we study.

The article is organized as follows. We first discuss the particle
properties of $\theta$. In Section \ref{evolution_sec}, we work
out the cosmological evolution of the fields, focusing in thermal and
 non-thermal production, and discuss
 the relic density of PGBs. The astrophysical and cosmological bounds, and the
consequences of the $\theta$ decay, are presented in Section \ref{astro}. A final
section discusses the conclusions of our work.

\section{$\theta$ - Mass And Lifetime}
\label{mass}

\begin{figure}[htb]
\begin{center}
\includegraphics[width=14cm, height=10cm]{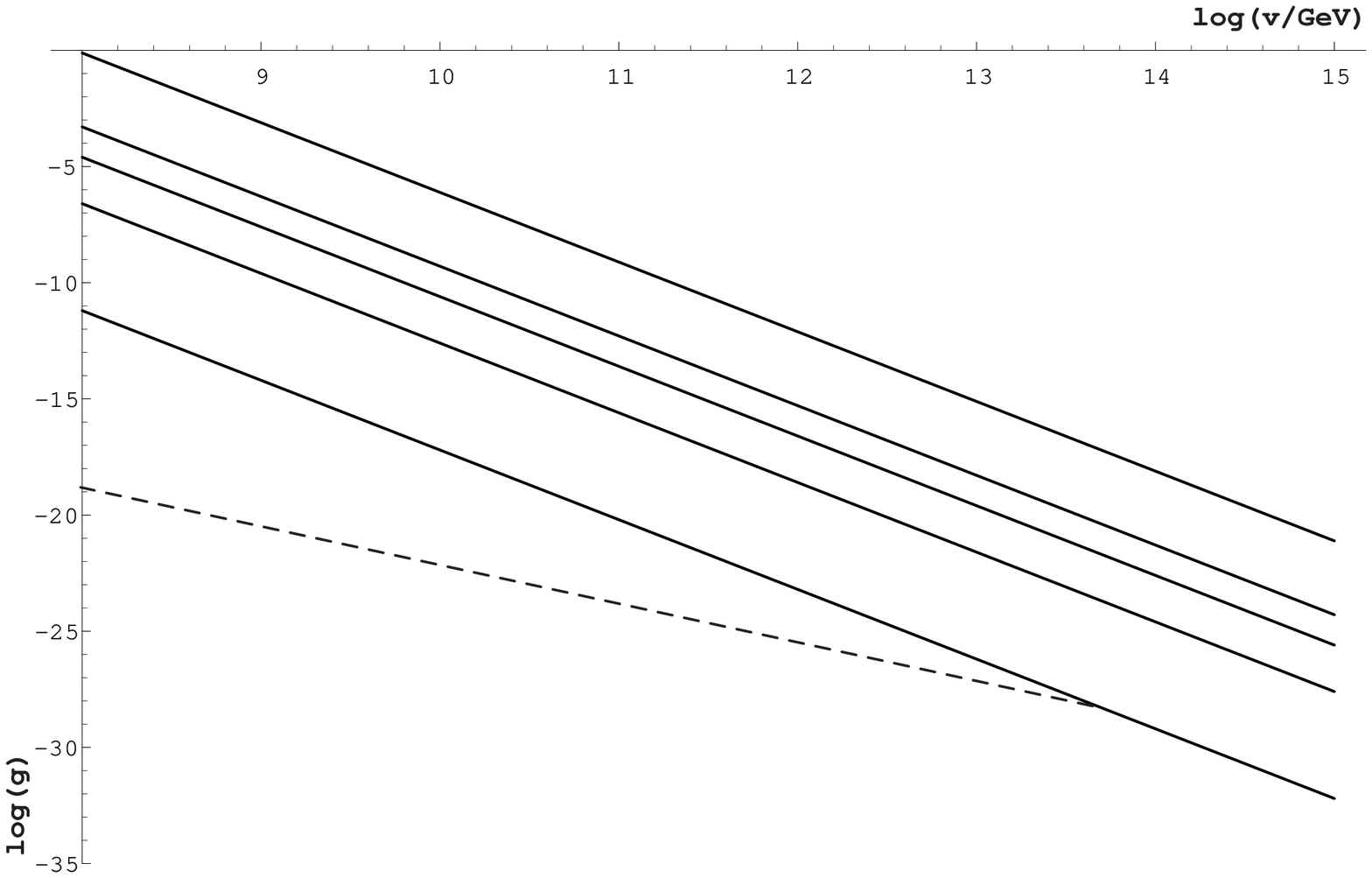}
\end{center}
\caption{\label{figA} From bottom to the top, the five solid lines
are the lines of constant mass, in the $g$,$v$ plane, for
$m_{\theta}=2m_e,\,2m_{\mu},\,2m_{p},\,2m_{b}$, and  $2m_t$. The
dashed line corresponds to the $\theta$ lifetime equal to the
universe lifetime, $\tau = t_0$, when $\theta
\rightarrow\gamma\gamma$ is the only available mode.}
\end{figure}

We have deduced the expression (\ref{thetamass}) that
gives the mass of the $\theta$-particle as a function of $v$ and
$g$. In terms of these two parameters, we plot in Fig.\ref{figA} lines of constant mass, for masses
corresponding to the thresholds of electron, muon, proton, bottom,
and top final decay,
$m_{\theta}=2m_e,\,2m_{\mu},\,2m_{p},\,2m_{b}$, and $2m_t$.

The lifetime $\tau$ of the $\theta$-particle depends on its mass
and on the effective couplings. A channel that is always open is $\theta
\rightarrow\gamma\gamma$. The corresponding width is
\begin{equation}
 \Gamma(\theta\rightarrow\gamma\gamma)=\frac{B_{\gamma\gamma}}{\tau}
=g^{2}_{\theta\gamma\gamma}\; \frac{m^{3}_{\theta}}{64\pi}
\end{equation}
where $B_{\gamma\gamma}$ is the branching ratio. When
$m_{\theta}>2m_f$, the decay $\theta\rightarrow f \bar{f}$ is
allowed and has a width
\begin{equation}
 \Gamma(\theta\rightarrow f \bar{f})
=\frac{B_{f\bar{f}}}{\tau} = g^{2}_{\theta f\bar f}\;
\frac{m_{\theta}}{8\pi}\,\beta
\end{equation}
where $\beta =\sqrt{1-4\,\left(\frac{m_f}{m_\theta}\right)^2}$.

For masses $m_{\theta}<2m_e$ the only available decay is into
photons. When we move to higher
$\theta$ masses, as soon as $m_{\theta}>2m_e$, we have the decay
$\theta\rightarrow e^+ e^-$. Actually, then the $\theta\rightarrow
e^+ e^-$ channel dominates the decay because
$\theta\rightarrow\gamma\gamma$ goes through one loop. However,
$\Gamma(\theta\rightarrow\gamma\gamma) \propto m_\theta^3$ whereas
$\Gamma(\theta\rightarrow f\bar{f}) \propto m_\theta$. By increasing
$m_\theta$ we reach the value $m_\theta \simeq 150\, m_e$, where
$\Gamma(\theta\rightarrow\gamma\gamma) \simeq
\Gamma(\theta\rightarrow e^+\,e^-)$. For higher $m_\theta$ the
channel $\theta\rightarrow\gamma\gamma$ dominates again. If we continue increasing the
$\theta$ mass and cross the threshold $2m_{\mu}$, the decay
$\theta\rightarrow \mu^+\mu^-$ dominates over $\theta\rightarrow
e^+ e^-$ and $\theta\rightarrow\gamma\gamma$, because $g_{\theta f\bar f}$ is proportional to the fermion mass
$m_f$. This would be true
until $m_\theta \simeq 150 \,m_\mu$, but before, the channel $\theta\rightarrow p\bar{p}$ opens, and so
on. For larger masses, each time a threshold
$2m_f$ opens up, $\theta\rightarrow f \bar{f}$ happens to be the
dominant decay mode.

\begin{figure}[htb]
\begin{center}
\includegraphics[width=14cm, height=10cm]{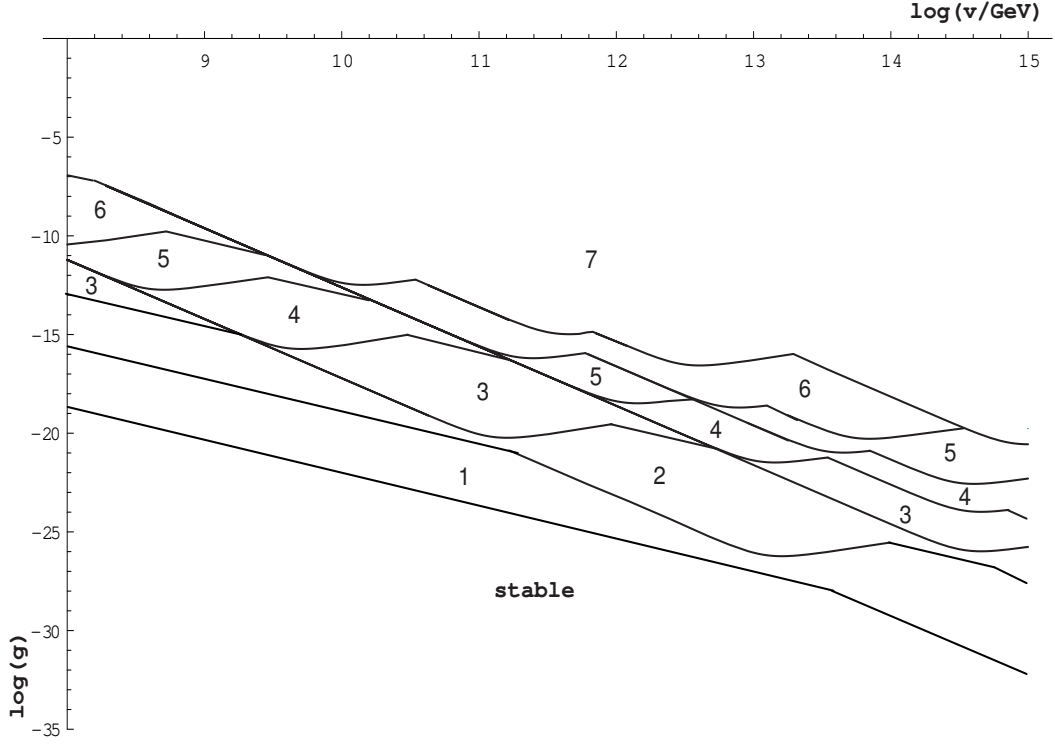}
\end{center}
\caption{\label{figB} Zones corresponding to some ranges for the
$\theta$ lifetime, $\tau$. At the bottom, we have the region of
the parameter space leading to a stable $\theta$ ($\tau>t_0$), and
at the top, the zone 7, where $\tau<1$ s. The other regions
correspond to : $(1)\, t_0 >\tau>10^{13}\,\rm s$;  $(2)\,
10^{13}\,\rm s>\tau>10^{9}\,\rm s$;  $(3)\, 10^{9}\,\rm
s>\tau>10^{6}\,\rm s$; $(4)\, 10^{6}\,\rm s>\tau>10^{4}\,\rm s$;
$(5)\, 10^{4}\,\rm s>\tau>300\,\rm s$ and $(6)\, 300\,\rm
s>\tau>1\,\rm s$.}
\end{figure}

Now we can identify in Fig.\ref{figB} the regions with different
lifetimes. We start with the stability region $\tau>t_0 \simeq 4\times 10^{17}$ s \cite{wmap}, the universe lifetime,
a crucial region for the dark matter issue. When
$m_{\theta}<2m_e$, $\tau = t_0$
along the dashed line in Fig.\ref{figA} until $v\simeq 4\times 10^{13}$ GeV and along the line $m_{\theta} = 2m_e$ for higher $v$. Below these lines, relic $\theta$ particles
would have survived until now; in practical terms, for values of
$g$ and $v$ in that region, we consider $\theta$ as a stable
particle. Above the lines, $\tau<t_0$ and the particle is
unstable. For future use, we define the
ranges: $(1)\, t_0
>\tau>10^{13}\,\rm s$;  $(2)\, 10^{13}\,\rm s>\tau>10^{9}\,\rm s$;
$(3)\, 10^{9}\,\rm s>\tau>10^{6}\,\rm s$;  $(4)\, 10^{6}\,\rm
s>\tau>10^{4}\,\rm s$;  $(5)\, 10^{4}\,\rm s>\tau>300\,\rm s$; and
$(6)\, 300\,\rm s>\tau>1\,\rm s$.

\section{Cosmological production and PGB density}
\label{evolution_sec}

$\theta$-particles can be produced in the early universe by
different mechanisms. There are non-thermal ones, like production
associated with the $\theta$ field oscillations, and production
coming from the decay of cosmic strings produced in the SSB. Also,
there could be thermal production. In this section we consider
these production mechanisms and the PGB density resulting from
them.

Let us begin with the $\theta$ field oscillation production. The expansion of the universe is characterized by the Hubble expansion rate $H$,
\begin{equation}
H =\frac{1}{2t} =  \sqrt{\frac{4 \pi^3}{45}}\, \sqrt{g_\star}\,
\frac{T^2}{M_P} \label{H}
\end{equation}
These relations between $H$ and the time and temperature of the
universe, $t$ and $T$, are valid in the radiation era. For the
period of interest, we take the relativistic degrees of freedom
$g_\star = 106.75$ \cite{KolbTurner}. In the evolution of the
universe, phase transitions occur when a symmetry is broken. When
the universe cools down from high temperatures there is a moment
when the potential starts changing its shape, and will have
displaced minima. This happens around a critical temperature $T_{
cr}\sim v$ and a corresponding critical time $t_{cr}$.

In our model, the explicit symmetry breaking is very small and the
evolution equations of $\rho$ and $\theta$ are approximately
decoupled. The temporal development leading $\rho$ and $\theta$ to
the minimum occurs in different time scales, since the gradient in
the radial direction is much greater than that in the angular one.
The evolution happens in a two-step process. First $\rho$ goes
very quickly towards the value $<\rho>=0$, and oscillates around
it with decreasing amplitude (particle decay and the expansion of
the universe work as a friction). Shortly after, $\rho$ will be
practically at its minimum. In this first step where $\rho$
evolves, $\theta$ does not practically change, maintaining its
initial value $\theta_0$, which in general is not the minimum,
i.e., $\theta_0 \neq 0$. $\theta$ evolves towards the minimum once
the first step is completely over. This is the second step where
the field $\theta$ oscillates around $<\theta>=0$.

We have confirmed these claims with a complete numerical
treatment of the full evolution equations, using the effective potential taking
into account finite temperature effects. We summarize our results in Appendix
\ref{appendixA}.

Hence, we can write the following $\rho$-independent evolution equation for $\theta$
\begin{equation}
\ddot{\theta}(t)+3H\, \dot{\theta}(t)+ m_\theta^2\, \theta =0
\label{evolution}
\end{equation}
Here overdot means $d/dt$. We do not introduce a
decay term in the motion equation (\ref{evolution}) and will treat the $\theta$ decay in the usual way. We can do it because $\theta$ has a lifetime $\tau$ that is much greater than the period of oscillations.

Oscillations of the $\theta$ field start at a temperature $T_{osc}$ such that $3H(T_{osc}) \sim m_\theta$  and this will be much later than the period of $\rho$ evolution provided $T_{osc}\ll T_{cr}\sim v$. This is equivalent to
\begin{equation}
g^{1/2}\, \ll\,3\times 10^{-3}\, \left( \frac{10^{11}\, \GeV}{v}
\right)^{-1/2} \label{ineq_g}
\end{equation}
We shall see that the phenomenologically viable and interesting
values of $g$ are very tiny, so that we can assume that the
inequality (\ref{ineq_g}) is fulfilled.

Since the $\theta$ oscillations are decoupled from $\rho$ oscillations, $\theta$ production is equivalent to the misalignment production for the axion and we can use those results \cite{three_papers}. The equation of motion (\ref{evolution}) leads to
coherent field oscillations that correspond to non-relativistic
matter and the coherent field energy corresponds to a condensate of
non-relativistic $\theta$ particles.

We define $\rho_{osc}$ and $n_{osc}$ as the energy density, and number density respectively, of the PGBs coming from the $\theta$ oscillations. The energy stored initially is
\begin{equation}
\rho_{osc} = m_\theta\,n_{osc} \simeq \frac{1}{2}\, m_\theta^2 \theta_0^2
\end{equation}
The initial angle is unknown, $\theta_0/v \in [0,2\pi)$, so that
we expect it to be of order one, $\theta_0/v\sim 1$. In
the following we will set $\theta_0= v$. Barring an unnatural $\theta_0$ fine tuned extremely close to 0, other choices of $\theta_0$ would lead essentially to the same conclusions we reach.

Let us now consider production of PGBs by cosmic string decays.
When our $U(1)$ symmetry is spontaneously broken, a network of
(global) strings is formed \cite{vilenkin}. These strings evolve
in the expanding universe and finally decay into PGBs, at a time
$t_{str}$ given by $m_{\theta}t_{str}\sim 1$, which is of the
order of $t_{osc}$. The same issue has been extensively studied
for the axions and we can borrow the results \cite{Davis,Harari}.
Unfortunately, there are different calculations that do not agree
among them. To be conservative, we take the least restrictive
result, namely, the one giving less particle production. That is
\cite{Harari}
\begin{equation}
n_{str}(t_{str})\approx\frac{v^2}{t_{str}}
\end{equation}
This is of the same order of magnitude as the PGB density at time $t_{str}$
due to $\theta$-oscillation production.

Thus, the total number density of non-thermally produced PGBs is
\begin{equation}
  n_{non-th} = n_{osc} + n_{str}
  \label{densit_no_therm}
\end{equation}

Let us now consider the thermal production of PGBs. Any species
that couples to the particles present in the early universe and
has a production rate $\Gamma$ larger than the Hubble expansion
rate $H$ during a certain period will be thermally produced.
Whether such a period exists or not has been investigated for the
axion \cite{Turner:1986tb,MRZ}. We adapt the axion results, taking
into account the different magnitude of the gluon-gluon vertex,
and conclude that for
\begin{equation}
\label{v_th}
v<7.2\times 10^{12}\,{\rm GeV} \equiv v_{th}
\end{equation}
there is always thermal production of $\theta$ in the early universe. However, for larger values of $v$, $\theta$ interacts so
weakly that $\Gamma<H$ always, or $\Gamma>H$ only for such a brief period of time that in practice there is no thermal production. We
denote this region by (I) in Fig.\ref{figC}.

When $v< v_{th}$, the $\theta$ species actually interacts with the
plasma in the following range of temperature $T$
\begin{equation}
v\geq T\geq \frac{v^2}{1.8\times 10^{14}\GeV}\equiv T_{end}
\label{thermrange}
\end{equation}
and a thermal population of $\theta$ is created with a number
density
\begin{equation}
n_{th}=\frac{\zeta(3)}{\pi^2}\,T^3 \simeq 0.12\,T^3
\label{thermdens}
\end{equation}

We should now reconsider the fate of the non-thermal PGBs in
(\ref{densit_no_therm}), since a thermalization period
(\ref{thermrange}) may be at work. For this discussion we shall
use the fact that $\theta$-oscillations and string decay start at
about the same temperature, $T_{osc}\sim T_{str}$. Let denote this
common temperature where non-thermal production starts by
$T_{\ast}$. When $v< v_{th}$, we have to distinguish between two
possible cases. First, if $T_{\ast}<T_{end}$, non-thermal $\theta$
are born after the thermalization period is over. We end up having
both a thermal $n_{th}$ and a non-thermal $n_{non-th}$ densities,
and a total density given by the sum $n_\theta=n_{th}+n_{non-th}$.
This first case corresponds to values of $v$ and $g$ indicated in
Fig.\ref{figC} by regions (II) and (III); in (II)
$n_{non-th}>n_{th}$, while in (III) $n_{th}>n_{non-th}$. The
second case corresponds to having $T_{\ast}>T_{end}$; if this is
the case, non-thermal PGBs will be in contact with the thermal
bath and will consequently thermalize. Indeed, independently of
details, we end up the period corresponding to (\ref{thermrange})
with $n_\theta=n_{th}$ given in (\ref{thermdens}). The region (IV)
in Fig.(\ref{figC}) is where such complete thermalization happens.

A word of caution is now in order. When $T_{\ast}>v$, radial and
angular oscillations are not decoupled. The analysis is not as
simple as we presented before: the oscillations cannot be
approximated by harmonic ones, and depend on initial conditions.
However, if $v<v_{th}$, the PGBs get thermalized in any case, so
that we have not to worry about the non-thermal production
details. For $v>v_{th}$ (and still $T_{\ast}>v$), since there is
no thermalization, the final density is $n_{\theta}=n_{non-th}$.
Admittedly, we have no general formula for the density in this
case, represented by (V) in Fig.\ref{figC}. In this region,
$\theta$ lifetimes are so small that the particle does not play
any cosmological role at all.

Summarizing, a number density of $\theta$ particles always appears
in the early universe. In regions (II), (III), and (IV), the
thermal density is $n_{th}(T=T_D)$ at decoupling temperature $T_D$
(the temperature where $\Gamma = H$). In regions (I), (II), and
(III) there is also a non-thermal  density $n_{non-th}$ at
formation temperature $T=T_{\ast}$.

As usual, the expansion of the universe dilutes these densities.
Eventually, if the particle is unstable it will decay. The issue
of unstable $\theta$ and effects of the decay will be worked out
in Section \ref{astro}, and here we focus on stable $\theta$
particles, since certainly they would be dark matter. To calculate
the relic density of $\theta$ today, apart from the expansion
effect, one has to take into account the transferred entropy to
the thermal bath, due to particle-antiparticle annihilations. We
follow the standard procedure (see \cite{KolbTurner}), and find
that
 the number density today is
\begin{equation}
n(T_0)=\frac{g_{\star s}(T_0)}{g_{\star s}(T_1)}\left(\frac{T_0}{T_1}\right)^3
n(T_1)
\end{equation}
where $T_1=T_D$ for $n_{th}$ and $T_1=T_{\ast}$ for $n_{non-th}$,
and $T_0= 2.7K$  is the photon temperature today. We take $g_{\star s}(T_1)=106.75$ and $g_{\star s}(T_0)=43/11$.

With all that, we can calculate the cosmological density of
$\theta$ particles that we would have today, $n_{\theta}(t_0)$,
and the corresponding energy density normalized to the critical
density $\rho_c$,
\begin{equation}
\Omega_{\theta}=\frac{m_{\theta}\,n_\theta(t_0)}{\rho_c}
\label{omega}
\end{equation}
with $\rho_c=0.5\times 10^{-5}\,\GeV\, cm^{-3}$ \cite{PDG2002}.

\begin{figure}[htb]
\begin{center}
\includegraphics[width=14cm, height=10cm]{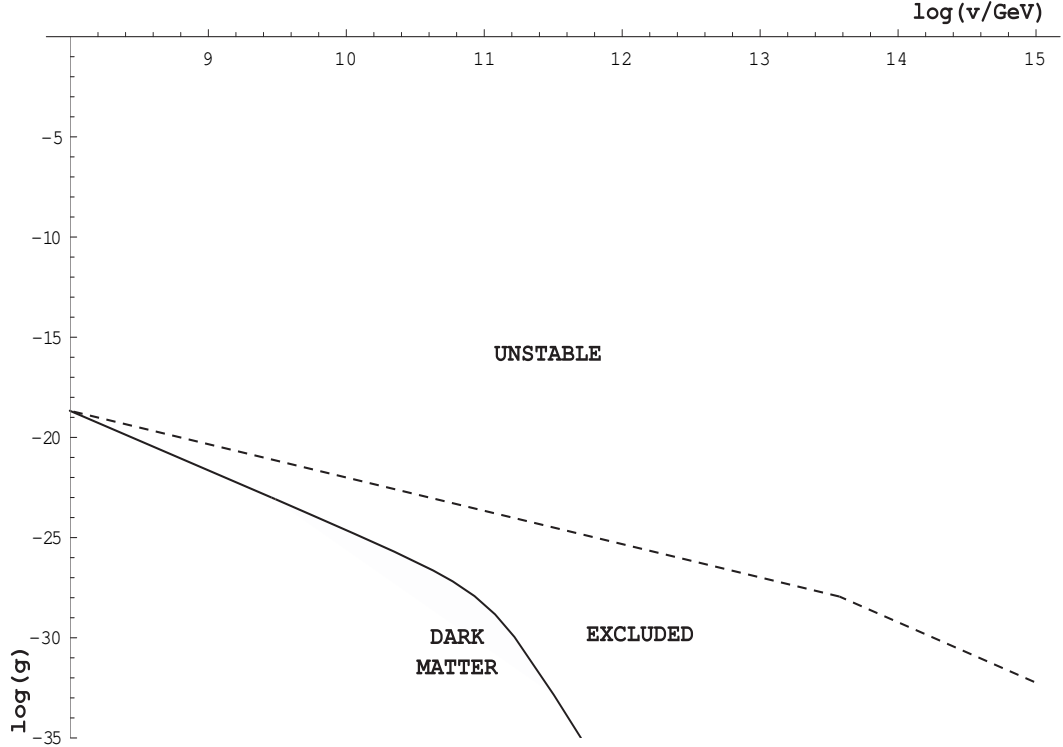}
\end{center}
\caption{\label{figC_p} Properties of $\theta$ before considering
the astrophysical constraints and the effects of the decay. Above
the dashed line of the plot we have the region of $g,\,v$ where
$\theta$ is unstable. The solid line corresponds to $\Omega_\theta
= 0.3$. The interesting dark matter region is labeled in the plot.
The region between the solid and dashed lines is excluded because
$\theta$ is stable and $\Omega_\theta>0.3$.}
\end{figure}

All cosmological data available lead to a dark matter contribution
$\Omega_{DM}\simeq0.3$ \cite{wmap}. In Fig.\ref{figC_p} we plot
the line $\Omega_{\theta}=0.3$. We will see in Section \ref{astro}
that part of this line is excluded. Clearly, values of $g$ and $v$
along the non-excluded part of the line would imply the density
needed to fit the cosmological observations. Our $\theta$ particle
is a dark matter candidate provided $g$ and $v$ are on the line,
or not far, in such a way that $\Omega_{\theta}$ is still a
substantial fraction of 0.3. We stress that in the region of
interest for dark matter it is the non-thermal production that
dominates (see Fig.\ref{figC}). Thus, the produced PGBs are
non-relativistic and would be cold dark matter. Values for
$\Omega_{\theta}$ much smaller than $\sim 0.3$ are allowed but not
cosmologically interesting. On the other side of the line, but
still in the stability region, we
 have the excluded region $\Omega_{\theta}>0.3$.

\section{Astrophysical and cosmological Constraints}
\label{astro}

\begin{figure}[htb]
\begin{center}
\includegraphics[width=14cm, height=10cm]{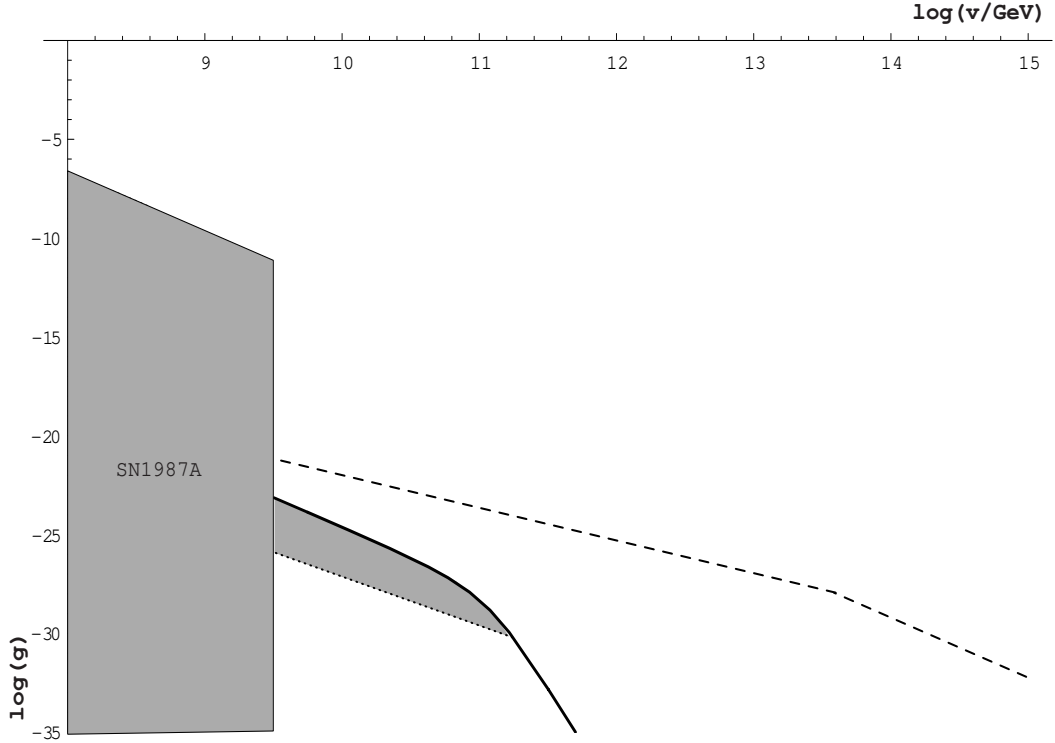}
\end{center}
\caption{\label{figD} The astrophysical constraints forbid the
shadow region labeled SN1987A. The other shadow region
corresponds to stable particles ($\tau>t_0$), yet the contribution
to the photon background is too high and thus this region is
forbidden. To help the reader, we also plot the line
$\Omega_{\theta}=0.3$ (solid) and the border between stable and
unstable PGBs (dashed).}
\end{figure}

PGBs would be emitted from the hot stellar cores since nucleons,
electrons and photons initiate reactions where $\theta$ is
produced. Provided $\theta$ escapes the star, the emission leads
to a novel energy loss channel, which is constrained by stellar
evolution observations using what is nowadays a standard argument
\cite{Raffelt_book}. The limits we find for $v$ are similar to the
limits found for the Peccei-Quinn scale \cite{PDG2002} because the
coupling of $\theta$ to matter is the same of the axion. We get
the bound
\begin{equation}
v\gtrsim 3.3\times 10^9\, \GeV \label{limvnucl}
\end{equation}
that is shown in Fig.\ref{figD}.

We see that the astrophysical constraint is on $v$, but not on
$g$, so that we may have relatively massive PGBs, while in the
axion case this leads to $m_a\lesssim O(10^{-3}{\rm eV})$, i.e.,
to forbid the existence of relatively massive axions. This makes
the PGB in our model an unstable particle, at variance with the
invisible axion.

Let us now investigate the region where $\theta$ is unstable, see
Fig.\ref{figC_p}. The cosmological effects of the decay will
depend on the lifetime $\tau$,
mass $m_\theta$, and number density $n_{\theta}$ at the moment of
the decay. These properties depend on the parameters $g$
and $v$. Observational data will help us to constrain different
regions of the $g$, $v$ parameter space. We will use several
pieces of data, depending on when the $\theta$ decay occurs. To do this systematically, it is convenient to distinguish
between the following time ranges, plotted in Fig.\ref{figB}
\begin{xalignat*}{3}
&{\rm range\,1}:&t_{0} &\quad >\; t\; > \quad t_{\rm rec} \simeq 10^{13} \,\rm s \\
&{\rm range\,2}:&t_{\rm rec}   &\quad >\;t\;>\quad t_{\rm c} \simeq 10^9 \,\rm s \\
&{\rm range\,3}:&t_{\rm c}     &\quad >\;t\;>\quad t_{\rm th} \simeq 10^6 \,\rm s \\
&{\rm range\,4}:&t_{\rm th}    &\quad >\;t\;>\quad t_{\rm He} \simeq 10^4 \,\rm s \\
&{\rm range\,5}:&t_{\rm He}    &\quad >\;t\;>\quad t_{\rm endBBN} \simeq 300 \,\rm s \\
&{\rm range\,6}:&t_{\rm endBBN}&\quad >\;t\;>\quad t_{\rm BBN} \simeq 1 \,\rm s\end{xalignat*}

We do not consider times earlier than $t_{\rm BBN}$ since there
are no observational data that give us any constraint at that times. The physical meaning of the chosen
times are
\begin{itemize}
\item[-]$t_{\rm rec}$ is the time of recombination, i.e., the
moment at which photons last scattered with matter, when free
electrons present in the cosmic plasma were bounded into
atoms;
\item[-]$t_{\rm c}$ is the time at which the
rate of Compton scattering $e\gamma \rightarrow e\gamma$ becomes
too slow to keep kinetic equilibrium between photons and
electrons;
\item[-]$t_{\rm th}$ is the time at which the
double-Compton scattering $e\gamma\rightarrow e\gamma\gamma$
decouples;
\item[-]$t_{\rm He}$ is the time at which the energy of
the Cosmic Background Radiation (CBR) is low enough to permit
photons of energy $\sim 10$ MeV to photodissociate $^4$He nuclei;
\item[-]$t_{\rm endBBN}$ is the time at which the Big Bang
Nucleosynthesis (BBN) finishes;
\item[-]$t_{\rm BBN}$ is the time at which BBN begins.
\end{itemize}

We can constrain the parameters $v$ and $g$ because the photon
spectrum would be distorted by $\theta$ decay products, and the
abundances of the light elements would be altered. We will discuss
this in the next two subsections.

\subsection*{Effects on the photon spectrum}

The products of the $\theta$ decay can be either photons or
fermions. When the decay products are fermions it is important to
note that if the number density of the created fermions is higher
than the photon density of the CBR, these fermions immediately
annihilate into photons, since we are considering decays produced
after the $e^+e^-$ annihilation in the early universe. The
distinction between the two channels will be important only when
the $\theta$-decay occurs in the time ranges 1 and 2, since in the
other regions photons and fermions thermalize.

When $\tau$ is in the range 1, it is convenient to distinguish
between two regions (see Figs.\ref{figA} and \ref{figB}). In the
region where $m_\theta<2\,m_e$, $\theta$ decays only into photons. The products of the other region are
mainly $e^-$ and $e^+$, but the values of
$v$ and $g$ in this region are such that the number density of the
created fermions is higher than the CBR photon density. Then, in
all the time range 1, the final products are photons, but for
$m_\theta>2\,m_e$ we use $\tau$ for the $\theta\rightarrow
e^+e^-$ decay. Photons produced in this range stream freely and
contribute to the diffuse photon background of the universe. One
can compute the present energy flux of photons per energy and
solid angle interval coming from $\theta$-decay
\cite{masso&toldra}
\begin{equation}
\frac{d^2\mathcal{F}_E}{dEd\Omega}=\frac1{2\pi}\,
\frac{n_{\theta}(t_0)}{\tau\,H_0}\left(\frac{E}{m_{\theta}/2}\right)^{3/2}\exp{\left[-\frac23\,\frac{1}{\tau\,
H_0}\left(\frac{E}{m_{\theta}/2} \right)^{3/2}\right]}
\label{flux}
\end{equation}
where $H_0$ is the Hubble constant today and $n_{\theta}(t_0)$ is
the number density of $\theta$ that we would have today if it had
not decayed. The predicted flux (\ref{flux}) is bounded by the
observational limits \cite{Ressell&Turner} on the photon
background flux in the energy range of interest. This restriction
excludes all the region in the $v$, $g$ parameter space
corresponding to time range 1. Taking into account that even for
$\tau > t_0$ a small fraction of PGBs have decayed into photons,
we can also prohibit part of the parameter space that corresponds
to stable particles with $\Omega\lesssim 0.3$ (see Section
\ref{evolution_sec}). In Fig. \ref{figD} we plot this excluded region.

In range 2, Compton scattering is not effective in maintaining
kinetic equilibrium between the $e^-$ and the $\gamma$ of the CBR.
Depending on the
precise values of $g$ and $v$, $\theta\rightarrow e^+e^-$,
$\theta\rightarrow\mu^+\mu^-$ or $\theta\rightarrow\gamma \gamma$
will be the dominant decay. The induced charged leptons, carrying
large energy, outnumber the existing CBR electrons. Different
processes now compete; one is scattering of these hot electrons
and muons with CBR photons. Another is $e^+e^-$ and $\mu^+\mu^-$
annihilations that give high energy photons, which heat CBR
electrons. The last is high energy photons produced in the decay
$\theta\rightarrow\gamma \gamma$, that also scatter and heat CBR
electrons which in turn scatter with CBR photons. Whatever
process is more important (it depends on the $g$, $v$ values), the
result is a distortion of the photon spectrum, parameterized by
the Sunyaev-Zeldovich parameter $y$ \cite{S-Z}. The energy $\Delta
E$ dumped to the CBR, relative to the energy of the CBR itself is
constrained by data on CBR spectrum \cite{PDG2002}
\begin{equation}
\frac{\Delta E}{E_{CBR}} \simeq 4y\lesssim 4.8 \times 10^{-5}
\label{SZ}
\end{equation}
where $\Delta E = m_\theta\,n_\theta$. The experimental bound (\ref{SZ}) rules out all
the $g$ and $v$ values that would imply a lifetime $\tau$ in the
time range 2.

In the range 3, Compton scattering is fast enough to thermalize
the products of the $\theta$-decay occurring in this range. This
is because even in  the region where the products are neither
photons nor $e^+\,e^-$, the
final particles will be photons in any case. In this region the
$e\gamma \rightarrow e\gamma\gamma$ processes are not effective so
the photon number cannot be changed. So, after
thermalization, one obtains a Bose-Einstein CBR spectrum with a
chemical potential, $f=\left[\exp{(E+\mu)/T}-1\right]^{-1}$,
instead of a black-body spectrum. The relation between $\mu$ and
$\Delta E$ is \cite{Toldra}
\begin{equation}
\left( \frac{4}{3}\,\frac{\zeta(2)}{\zeta(3)} -
\frac{\zeta(3)}{\zeta(4)}\right) \,\mu = \frac{\Delta E}{E_{CBR}}
- \frac{4}{3}\,\frac{\Delta n_\gamma}{n_\gamma} \label{mu_eq}
\end{equation}
($\zeta(n)$ is Riemann's zeta function). The parameter $\mu$ is very well
constrained by CBR data \cite{PDG2002} that gives $|\mu|< 9\times
10^{-5}$. This value rules out all the $g$, $v$ region that would
give $\tau$ in the time range 3.

For times before the range 3, both Compton and double Compton
scattering are effective, so the decay products thermalize with
the CBR, without disturbing the black body distribution but
changing the evolution of the temperature of the thermal bath.
This temperature  variation leads  to a change in the
photon number, and thus to a decrease on the parameter $\eta
\equiv n_b/n_\gamma$. The knowledge we have on the value of this
parameter at $t_{\rm rec}$ \cite{wmap} and $t_{\rm endBBN}$
\cite{Sarkar} allows to constrain the $g$, $v$ region  that would
give lifetimes $\tau$ in the time ranges 4 and 5. Although the
corresponding restriction is quite poor (it is essentially $\Delta
E/E_{CBR}\lesssim 1$), yet it excludes the parameter space
corresponding to the ranges 4 and 5. However, constraints from the
effects of the $\theta$-decay on the light element abundances are
much more restrictive in these ranges, as we will examine in the
next subsection.

\subsection*{Effects on the abundances of the light elements}

The period of primordial nucleosynthesis is the earliest epoch
where we have observational information. Also, the theoretical
predictions of the primordial yields of light elements are robust.
The agreement with observation is considered one of the pillars of
modern cosmology. The $\theta$ decays, and the $\theta$ particle
itself, might modify the abundances of the light elements, which
implies restrictions on the $v$, $g$ parameters.

First, we consider how the decays of a PGB affect the abundances
of the light elements after they are synthesized, i.e., after
$t_{\rm endBBN}$. One of the consequences of the decay is the
production of electromagnetic showers in the radiation-dominated
plasma, initiated by the decay products. As a result, photons of
energy $\sim$10 MeV scatter and photodissociate light elements.
This scattering occurs after $t_{\rm He}$, because before $t_{\rm
He}$, the collision of these photons with the CBR is more probable
than with the light elements. In time range 4, the
photodissociated element is deuterium (if $m_\theta > 10$ MeV)
\cite{Sarkar} while in time ranges 3 and 2 there is helium
photodissociation (if $m_\theta > 100$ MeV), with the subsequent
production of deuterium. Observational data for the abundance of
deuterium constraint all these processes. When $\theta$ decays
into quarks which hadronize subsequently, hadronic showers can
also be produced (if $m_\theta \sim 1$ GeV). These processes
dissociate $^4$He before $t_{\rm He}$, overproducing deuterium and
lithium. All these constraints, that are summarized in
\cite{Sarkar}, exclude all the $g$ and $v$ values that give $\tau$
in the time ranges 2,3,4 and 5, provided mass conditions are
fulfilled for each case. However, values from ranges 2,3,4 and 5
that do not satisfy the proper mass restrictions, are anyway
excluded by the constraints we considered in  the former subsection.

The other effect on the light element abundances arises because
the $\theta$ particle would modify the BBN predictions. The
presence of $\theta$ and, more important, the presence of the
relativistic products of the $\theta$-decay, accelerate the
expansion rate of the universe in the relevant BBN period and
modify the synthesis of the light elements. The decay of the
$\theta$ boson  is also a source of entropy production, which
alters the temperature evolution of the universe. This changes the
number of photons (and the value of $\eta$) and produces an
earlier decoupling of neutrinos. Then the relation between $T_\nu$
and $T_\gamma$ is changed, with potential effects on the BBN
physics. It is important to note that this production of entropy
never rises the temperature of the universe \cite{Turner}, and
does not lead to several successive BBN periods. All these effects
in BBN have been implemented \cite{Turner2} in the usual code,
which allows to constrain the quantity $\Delta E/n_\gamma$. As a
result, some of the values of $g$, $v$ that would give $\tau$ in
the time range 6 are ruled out. The potential effects of hadronic
showers, that we have mentioned earlier, also would modify the BBN
results \cite{Sarkar}, allowing us to exclude part of the range 6,
but not all of it.

All the results exposed in this section are summarized in
Fig.\ref{figE}. There, we can see that after our systematic study,
it turns out that all the zone of parameters that lead to
lifetimes between about 1 s and $t_0$ is ruled out. Only a small
zone corresponding to range 6 is allowed.

\section{Discussion and Conclusions}
\label{conclusions}

Gravitational arguments suggest that global symmetries are
explicitly violated. We describe this violation using an effective
theory framework that introduces operators of order higher than
four, suppressed by inverse powers of the Planck mass. These
operators are considered as a perturbation to the (globally)
symmetric part of the potential.

The SSB of global symmetries with a small explicit breaking leads
to PGBs: Goldstone bosons that have acquired a mass. Equivalent to
the appearance of a mass for the boson, there is no longer an
infinity of degenerate vacua.

In this article we have studied the cosmology of the PGB $\theta$
that arises in a model containing a scalar field with a potential
that can be divided in a part that has a global non-anomalous
$U(1)$ symmetry and another part with gravitationally induced
terms that are $U(1)$ violating. In our analysis we let vary two
parameters of the model: the SSB breaking scale, $v$, and the
coupling $g$ of the Planck-induced term in the potential. We have
analyzed the evolution of the field towards the vacuum in a phase
transition in the early universe. It occurs in a two-step process:
first the radial part attains its minimum in a relatively short
time and second the angular part of the field starts oscillating
well after the first step is over. The $\theta$ field oscillations
correspond to non-relativistic matter. We have calculated the
density of $\theta$ particles born through this mechanism and also
through the decay of cosmic strings created at the SSB. There
might also be thermal production of PGBs in the early universe and
also there might be thermalization of PGBs produced non-thermally;
these issues have been fully analyzed in our article.

A variety of arguments constrain the parameters $v$ and $g$ of
$\theta$. There are astrophysical constraints coming from energy
loss arguments. There are also cosmological bounds. When the
particle is stable its density cannot be greater than about the
critical density, otherwise the predicted lifetime of the universe
would be too short. If the particle is unstable the decay products
may have a cosmological impact. We have watched out for effects of
the decay products on the CBR and on the cosmological density of
light elements.

\begin{figure}[htb]
\begin{center}
\includegraphics[width=14cm, height=10cm]{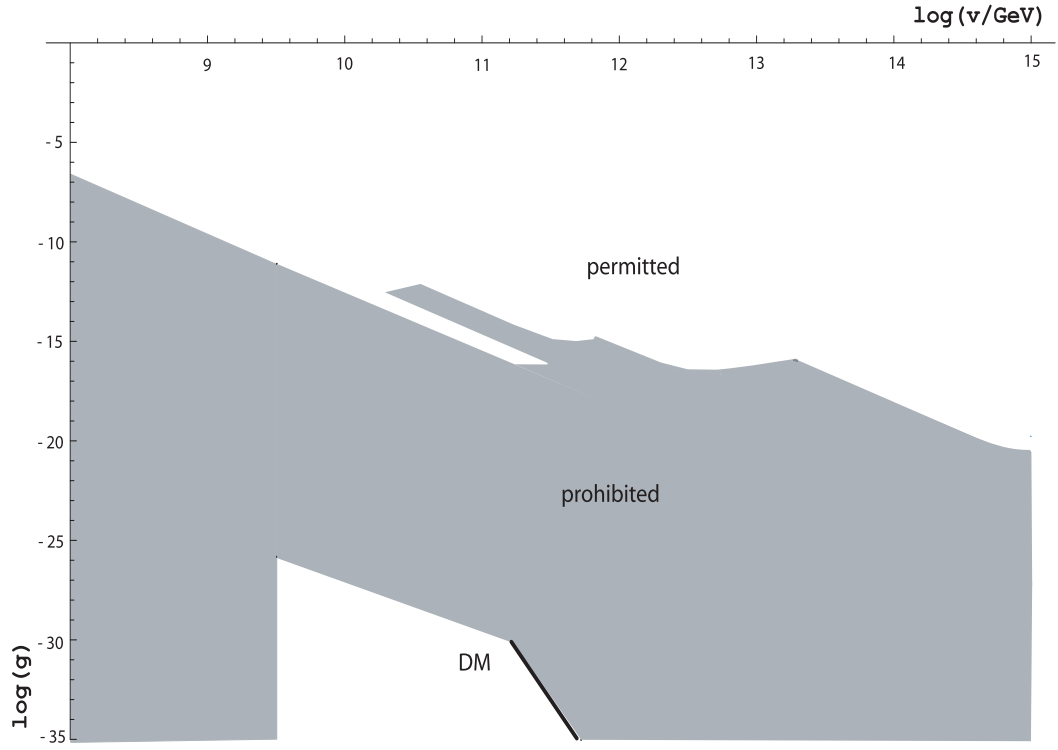}
\end{center}
\caption{\label{figE} The prohibited region, using all constraints
we have studied, is in shadow. In white, we show the allowed
region. The solid line corresponds to $\Omega_{\theta}=0.3$.}
\end{figure}

We have considered all the above potential effects and used
empirical data to put constraints on $g$ and $v$.
We have been led  to exclude the region
of the $g$, $v$ parameter space indicated in Fig.\ref{figE}.

In  Fig.\ref{figE}. we see that there are two allowed regions in
the plot. First, in the upper part of the plot there is an allowed
region.  It is where $\tau < $ 1 s, except the tooth at values that
are about $v\sim 10^{11}$ GeV and $g \sim 10^{-13}$ that
corresponds to 1 s $< \tau < $ 300 s (part of zone 6 in Fig.\ref{figB}). For a $\theta$ that
has the parameters corresponding to this first region, it will
definitely be extremely difficult to detect the particle. Also, in
any case, it will have no cosmological relevance.

The second permitted zone of the figure is where $\theta$ is dark
matter, at the bottom of the plot. It would be an interesting cold
dark matter candidate provided the values of $g$ and $v$ are not
far from the solid line in Fig.\ref{figE}. There is an upper limit
to the mass $m_\theta$ in the allowed region where $\theta$ is a
dark matter candidate
\begin{equation}
m_\theta \, \lesssim\, 20\, \rm eV \label{mDM}
\end{equation}
A way to detect $\theta$ would be using the experiments that try
to detect axions which make use of the
two-photon coupling of the axion. Since a similar coupling to two
photons exists for the PGB, we would see a signal in those
experiments \cite{Masso:2002ip}. The detection techniques use
coherent conversion of the axion to photons, which implies that in
order that $\theta$ would be detected, we should have $m_\theta <
10^{-3}$ eV.

For $\theta$ be dark matter, we notice that the values of
$g$ have to be very small
\begin{equation}
g < 10^{-30}
\label{tiny}
\end{equation}
We do not conclude that these values are unrealistically small.
Without any knowledge of how gravity breaks global symmetries it
would be premature to argue for or against the order of magnitude
(\ref{tiny}). For example, in \cite{0009030}, Peccei elaborates
about the explicit gravity-induced breaking of the Peccei-Quinn
symmetry, and gives the idea that perhaps the finite size of a
black hole when acting on microscopic processes further suppresses
Planck-scale effects.

Apart from that, there is an easy way to get PGBs as dark matter
candidates for values of $g$ not as tiny as in  (\ref{tiny}).
Notice that to work out the most simple case,  we have considered
$n=4$ in Eq.(\ref{Vnosym}). It suffices to consider more general potentials
\begin{equation}
V_{non-sym}=-\tilde{g}\frac1{M_P^{n+m-4}}|\Psi|^n\left(\Psi^m
e^{-i\delta}\, +\, {\rm h.c.}\ \right) \label{Vnosym_general}
\end{equation}
with $n,m$ integers. In the present article we made $n=4$, $m=1$.
If we take greater values, we get a suppression of the symmetry
breaking term due to extra factors $v/M_P$ and we may allow values
for $\tilde{g}$ higher than the ones obtained for $g$. In order of
magnitude, for $v \sim 10^{11}$ GeV, we see that taking operators
of dimension $n+m=8$ or $9$ we have a PGB that is a dark matter
candidate, but now with $\tilde{g} \sim {\cal O} (1)$. In building
the model we should have a reason for having the order of the
operators starting at $n+m>5$. The standard way is to impose
additional discrete symmetries in the theory.

Finally, we would like to comment on having put $g'=1$ in
(\ref{ferm_coup}). We could, of course, maintain  $g'$ free,  even
we could let  $g'$ be different for each fermion, but, in our
opinion, the introduction of extra parameters would make the
physical implications of our model much less clear. This is why we
fixed $g'=1$, but now it is time to think what happens for other
values of $g'$.

The coupling $g'$ appears accompanying a factor proportional to
the mass of the fermion and inversely proportional to the energy
breaking scale $v$, as expected for real and pseudo Goldstone
bosons. When a fermion has a $U(1)$ charge, we have no reason to
expect that $g'$ is much different than ${\cal O} (1)$, but if a
fermion has vanishing $U(1)$ charge then the coupling of $\theta$
to this fermion may only go through loops, and consequently we
have a smaller $g'$. In this case, an important change concerns
the astrophysical bound. Since $g_{\theta NN}$ is smaller than
$m_N/v$, the bound from supernova is weakened and values of $v$
smaller than $3.3\times 10^{9}$ GeV would be allowed. Another
effect is that the bounds coming from the cosmological effects of
the $\theta$ decay are relaxed, since the lifetimes are longer
when $g'$ is smaller. However, this does not mean that part of the
prohibited region in Fig.\ref{figE} would be allowed. We have to take into account that non-thermal production is not altered when changing $g'$
and therefore the bound $\Omega_\theta <$ 0.3 leads to strong
restrictions in the $g,v$ parameter space. Also, $\theta$ thermal
production is suppressed with smaller $g'$.

\begin{figure}[htb]
\begin{center}
\includegraphics[width=14cm, height=10cm]{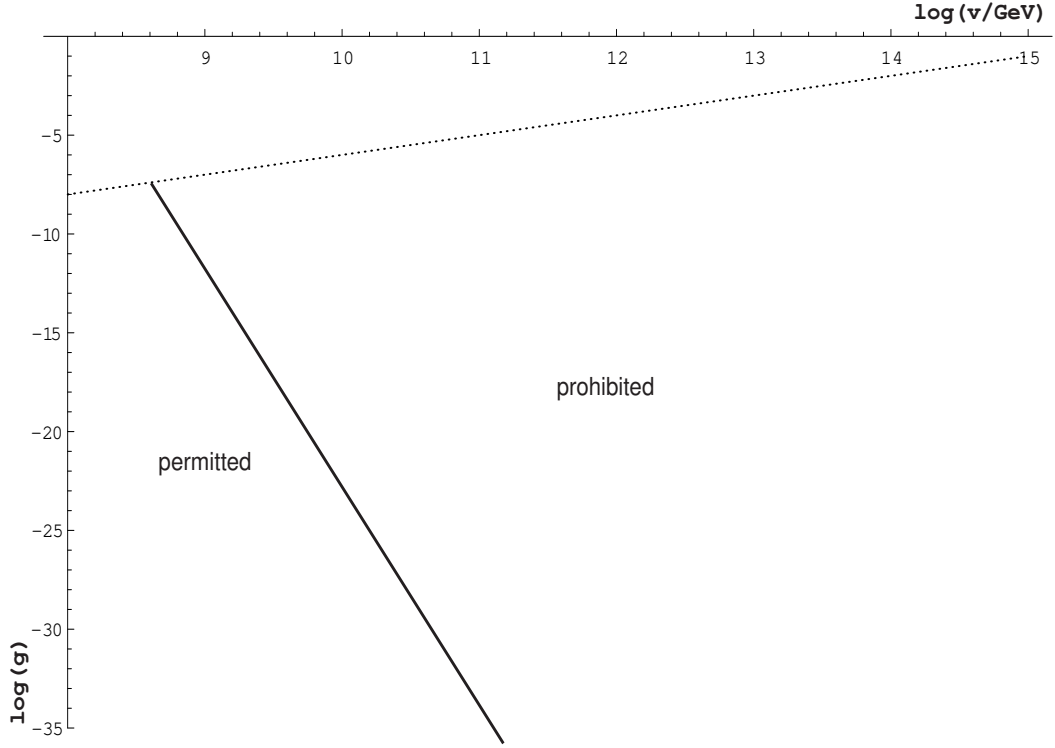}
\end{center}
\caption{\label{figE_p} Permitted and prohibited regions in the
limit $g'\rightarrow 0$. The solid line is $\Omega_\theta = 0.3$
and the dotted line is $T_{\ast}=v$. In the upper part of the
dotted line we have no reliable way to calculate $\Omega_\theta$.}
\end{figure}

We would like to show graphically what would happen for very small
$g'$, and with this objective we show in Fig.\ref{figE_p} the
permitted and the prohibited regions in the limit $g' \rightarrow
0$. At the view of the result, we conclude that one may have PGB
as a dark matter candidate for much larger values of $g$ than
obtained before in (\ref{tiny}).

\begin{acknowledgments}
We thank Mariano Quir{\'o}s and Ramon Toldr{\`a} for very useful
discussions. We acknowledge support by the CICYT Research Project
FPA2002-00648, by the EU network on Supersymmetry and the Early
Universe (HPRN-CT-2000-00152), and by the \textit{Departament
d'Universitats, Recerca i Societat de la Informaci{\'o}} (DURSI), Project
2001SGR00188. One of us (G.Z.) is supported by the DURSI, under grant 2003FI 00138.
\end{acknowledgments}

\appendix
\section{}
\label{appendixA}
\subsection{How to obtain the effective potential $V_{eff}$}
We present here in some detail how to find the effective potential
that gives us a complete description of the physics involved in
our model. Following the standard procedure \cite{Quiros}, taking
into account the finite temperature effects, we are led to a new
contribution to $V_{sym}$, which is given by
\begin{equation}
V^{\beta}=\frac{1}{2\pi^2\beta^4}J_{B}[m^2\beta^2]=
\frac{1}{2\pi^2\beta^4}\int_0^\infty\,dx\,x^2\ln\left[
1-e^{-\sqrt{x^2+\beta^2m^2}}\right]
\label{V_term}
\end{equation}
where $J_B$ is the thermal bosonic function and $\beta=1/T$, and
$m^2=-\frac12\lambda v^2+\lambda\Psi^{\star}\Psi+\frac12\lambda
T^2$ is the effective mass. With (\ref{V_term}), we see the
behavior of the finite temperature effective potential. For
practical applications, it is convenient to use a high temperature
expansion of $V^{\beta}$ \cite{Dolan_Jackiw} written in the form
\begin{equation}
V^{\beta}\simeq\frac1{24}m^2T^2-\frac1{12\pi}m^3T-\frac1{64\pi^2}
m^4\ln\frac{m^2/T^2}{223.63}
\end{equation}
where we have neglected terms independent of the field. The
effective potential must contain the explicit symmetry-breaking
term of our model, $V_{non-sym}$. Using for $\Psi$ the
parametrization $\Psi=\phi\,e^{i\theta/v}$, the expression for
this term is
\begin{equation}
V_{non-sym}=-2g\frac{\phi^{n+1}}{M_P^{n-3}}\cos\left(\frac{\theta}{v}
\right)
\end{equation}
Thus, our effective potential will be written as the sum of three terms
\begin{equation}
V_{eff}=V_{sym}+V^{\beta}+V_{non-sym}
\end{equation}

\subsection{How to find $T_{cr}$}
SSB of $V_{eff}$ is triggered at a critical temperature $T_{cr}$
corresponding to time $t_{cr}$, and instead of having a minimum at
$\phi=0$, a local maximum appears. $T_{cr}$ is the temperature
when the second derivative of $V_{eff}$ at $\phi=0$ cancels. In
calculating the derivative of $V_{eff}$, we find that
\begin{equation}
\frac{\partial V_{eff}}{\partial\phi}=\phi\cdot f(\phi,T)
\end{equation}
$f(\phi,T)$ contains all the information we need to find both
$T_{cr}$ and the minimum of the potential as a function of $T$
(writing $f(\phi,T)=0$ and solving for $\phi$). The second
derivative of $V_{eff}$ at its origin is
\begin{equation}
\left.\frac{\partial^2V_{eff}}{\partial\phi^2}\right|_{\phi=0}=
f(\phi,T)|_{\phi=0}+(\phi\,f'(\phi,T))|_{\phi=0}
=f(0,T)
\end{equation}
$T_{cr}$ is the solution to the equation $f(0,T)=0$, and depends
on $\lambda$ and $v$. In order of magnitude, $T_{cr}\sim v$.

\subsection{Field evolution}
At early times, the effective potential has a unique minimum at
$\phi=0$. As the temperature goes down, its effects on the
effective potential diminishes, and at $T_{cr}$ it is
spontaneously broken and a second order phase transition occurs.
In this subsection, we present our study on this phase transition
focusing on the separate evolutions of the radial ($\phi$) and
angular ($\theta$) parts of $\Psi$ and which are the implications
of $V_{non-sym}$ in our effective potential.

The evolution of the two fields mentioned above is described by
these equations
\begin{equation}\left\{\begin{array}{ccc}
 \ddot{\phi}(t)+(3H+\Gamma_{\phi}) \dot{\phi}(t)+\frac{\partial V_{eff}}
{\partial \phi} & = & 0\\
 \ddot{\theta}(t)+3H\dot{\theta}(t)+\frac{\partial V_{eff}}
{\partial\theta} & = & 0
\end{array}\right.
\label{eqs_set}
\end{equation}
where $H=1/(2t)$ is the Hubble expansion rate of the universe, and
overdot means time derivative. An important difference between the
two equations is the appearance of $\Gamma_{\phi}$ in the first
one, and this is because $\theta$ couples derivatively to matter,
see (\ref{ferm_coup}), while $\phi$ couples as $g_{\phi f\bar
f}\bar f f\phi$. Therefore, the coupling of $\theta$ is weak,
since it is suppressed by a high-energy scale $v$, but the
coupling of the radial part is not and has to be introduced in the
evolution equation for $\phi$. For our numerical simulations, we
put $\Gamma_{\phi}=m_{\phi}/8\pi=\sqrt{2\lambda}\,v/8\pi$, which
corresponds to $g´_{\phi f\bar f}=1$ and consider only one species
of fermions in the decay. The two differential equations
(\ref{eqs_set}) are not independent of each other. They are
related due to the fact that $V_{eff}$ has both $\phi$ and
$\theta$ dependence. However, the coupling between equations, in
the case where $g$ is tiny, is small and can be neglected in a
first approximation. This is what we have done in Section
\ref{evolution_sec}. Here we do not neglect it since we want to do
a complete numerical analysis and calculate the evolution when $g$
is not small and check when the scenario described in Section
\ref{evolution_sec} breaks down. It is convenient to replace the
temperature dependence of $V_{eff}$ with a time dependence, using
relation (\ref{H})
\begin{equation}
T^2=H\sqrt{\frac{45}{4\pi^3}}\sqrt{\frac1{g_{\ast}}}M_P=C\,t^{-1}
\end{equation}
Here $C=\frac1{34.3}M_P$ considering $g_{\ast}=106.75$ for the
temperature range where we apply the evolution equations. To
simplify calculations and graphical displays, we introduce new
dimensionless variables
\begin{equation}
\tilde{\phi}=\frac{\phi}{v};\hspace{1cm}\tilde{\theta}=\frac{\theta}{v};
\hspace{1cm}\tilde{t}=\frac{t}{t_{cr}};\hspace{1cm}t_{cr}=\frac{C}{T_{cr}^2}.
\label{tcrit}
\end{equation}
With these changes, the two evolution equations are
\begin{equation}\left\{\begin{array}{ccc}
\ddot{\tilde{\phi}}(\tilde{t})+(\frac3{2\tilde{t}}+\frac{\sqrt{2}}{8\pi}
\frac{C}{T_{cr}^2}\sqrt{\lambda}v) \dot{\tilde{\phi}}(\tilde{t})
+\frac{C2}{v^2T_{cr}^4}\frac{\partial V_{eff}}{\partial\tilde{\phi}} & =
& 0\\
 \ddot{\tilde{\theta}}(\tilde{t})+\frac3{2\tilde{t}}
\dot{\tilde{\theta}}(\tilde{t})+\frac{C2}{v^2T_{cr}^4}
\frac{\partial V_{eff}}{\partial\tilde{\theta}} & = & 0
\end{array}\right.
\label{evoleqs}
\end{equation}
We can numerically solve  the system (\ref{evoleqs}) for any
values of interest for $v$ and $g$. In Fig.\ref{figF}, we show one
such solution for arbitrarily chosen values for $v,g$, and
$\lambda$, for $n=4$.
\subsection{Discussion}
In Section \ref{evolution_sec} we based our work on the fact that,
due to the smallness of the explicit symmetry breaking, the field
evolution towards its minimum occurs in a two-step process:
firstly, the radial field goes quickly towards its vacuum
expectation value, oscillates around it for a finite time till it
stops due to the expansion of the universe and coupling to
fermions; secondly, the angular field stays constant much longer
and finally starts to oscillate around its minimum. We have proved
numerically that this is so; we have checked it by solving the
system equations (\ref{evoleqs}) for a variety of values of our
parameters.

We would like to specify the upper limit on $g$ for our model to
make sense. The parameter $g$ is considered to be too large when the term
$V_{non-sym}$ in the effective potential starts to dominate over
the other ones, for $\phi\sim v$. When this happens, the explicit
symmetry breaking is so big that, where the absolute minimum of
the effective potential is supposed to be, the first derivative of
$V_{eff}$ with respect to $\phi$ is negative and there is no
minimum at all. In this case, there makes no sense talking about
angular oscillations around the minimum. Therefore, by comparing
the $(\lambda/4)\phi^4$ term with $V_{non-sym}$, one obtains an
upper limit for $g$
\begin{equation}
g<\frac{\lambda}8\frac{M_P}{v}
\label{g_cond}
\end{equation}

\begin{figure}[htb]
\begin{center}
\includegraphics[width=14cm, height=10cm]{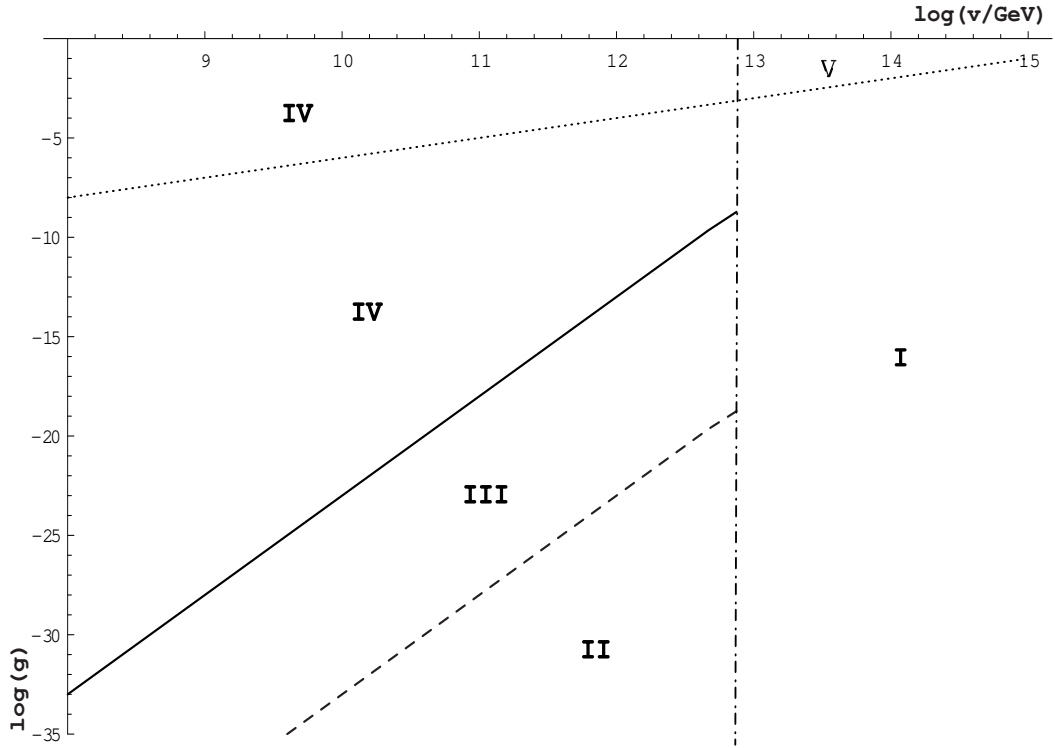}
\end{center}
\caption{\label{figC} We show the various regions according to the
$\theta$ production mechanism in the early universe. With this
objective, we display the lines: $T_{\ast}=v$ (dotted line),
$T_{\ast}=T_{end}$ (solid line), $n_{th}=n_{non-th}$ (dashed line)
and $v=v_{th}$ (dot-dashed line). In (I) and (V) there is only
non-thermal production. In (II) and (III) there are non-thermal
and thermal processes that generate $\theta$, in (II) non-thermal
dominates while in (III), it is thermal production that dominates.
In (IV) there is thermally and non-thermally produced $\theta$s,
but the last are thermalized. In (V), we have not a general
formula for predicting $n_{non-th}$.}
\end{figure}

\begin{figure}[]
  \centering
  \subfigure[]{
    \includegraphics[width=11.3cm, height=7cm]{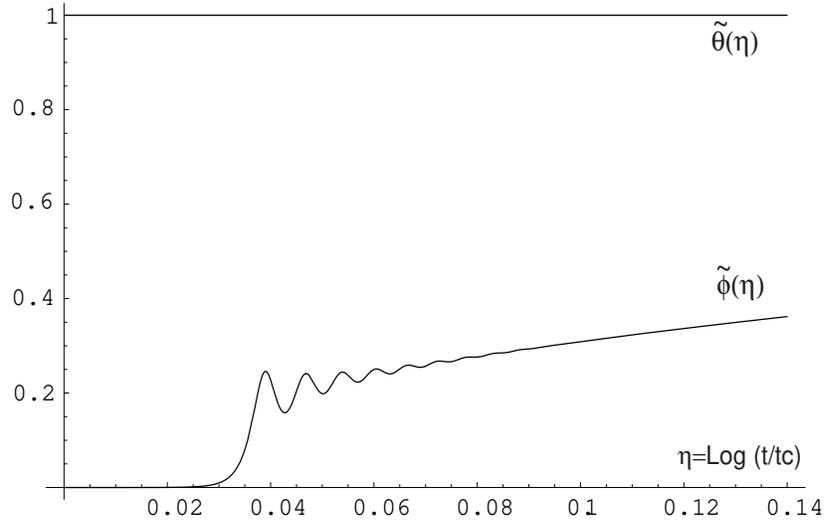}
    \label{fig1}}

  \subfigure[]{
    \includegraphics[width=11.3cm, height=7cm]{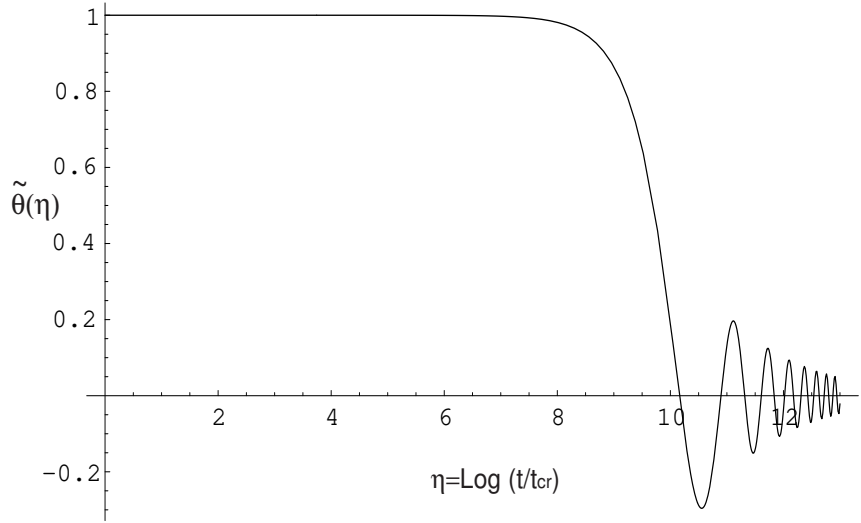}
    \label{fig2}}
\caption{Evolution of the normalized fields $\tilde \theta$ and $\tilde \phi$
as a function of $\eta=\log(t/t_{cr})$ with $t_{cr}$ defined in
(\ref{tcrit}). We see how $\tilde \phi$ evolves first while
$\tilde \theta$ remains constant (a), and how $\tilde \theta$
finally oscillates (b). Notice the (logarithmic) time scales. For
this numerical simulation we chose: $v=10^{11}$, $\lambda=10^{-2}$
and $g=10^{-8}$.} \label{figF}
\end{figure}

We have numerically checked that, for parameter values of interest
to us, radial oscillations start very early and they are very
rapidly damped, at a time scale much less than the time when
$\theta$-oscillations start. One example is plotted in
Fig.\ref{figF}. What happens is that the radial field oscillates
for a small time around the minimum of the symmetric part of the
effective potential and after the oscillations stop, the field
stays at the minimum and evolves in time until temperature effects
are irrelevant and the minimum stabilizes at
$\tilde{\phi}=1~(\phi=v)$. Thus, for values of $g$ that satisfy
(\ref{g_cond}) and (\ref{ineq_g}), when angular oscillations
start, the radial ones have already stopped and we must not worry
about the possibility that the two oscillations happened at  the
same time. More important is to impose the condition that when
radial oscillations start, the minimum of $V_{eff}$ be located
close to the value $\phi=v$ in order to have initial conditions
for $\theta$-oscillations also of order $v$. With all this in
mind, for $v$ values in the range of interest $(10^8$
GeV$<v<10^{15}$ GeV), we get to the conclusion that $g$ must be
smaller that about $10^{-5}$ (the number depends slightly on $v$
and $\lambda$). In particular, considering values of interest for
$\theta$ to be a dark matter candidate ($v\sim 10^{11}$ GeV) and
$\lambda \sim 10^{-2}$, we obtain an approximate limit
$g<10^{-5}$. This upper limit is also roughly given by the dotted
line represented in Fig.\ref{figC}, which corresponds to
$T_{\ast}\sim v$. Obviously, values for $g\sim 10^{-30}$ and
$v\sim 10^{11}$ GeV which we have found to be interesting to have
$\theta$ as a reasonable dark matter candidate of the universe,
are consistent with our mechanism of producing $\theta$ particles.




\newpage

\end{document}